\documentclass[adp,fleqn]{w-art}
\usepackage{times}
\usepackage{w-thm}
\usepackage[]{graphicx}
\chardef\bslash=`\\ 

\begin{document}
\DOIsuffix{theDOIsuffix}
\Volume{12}
\Issue{1}
\Copyrightissue{01}
\Month{01}
\Year{2003}
\pagespan{1}{}
\Receiveddate{} 
\Accepteddate{}



\title[FCS of 1D metals with impurities]{Statistics of charge transfer through impurities in
strongly correlated 1D metals}


\author[A.\ Komnik]{A. Komnik\inst{1,2}}
\address[\inst{1}]{Institut f\"ur Theoretische Physik, Universit\"at Heidelberg,
Philosophenweg 19, D--69120 Heidelberg, Germany} 
\address[\inst{2}]{Physikalisches Institut, Universit\"at Freiburg, Hermann-Herder-Str. 3,
D--79104 Freiburg, Germany} 
\author[B.\ Trauzettel]{B. Trauzettel\inst{3}}
\address[\inst{3}]{Departement Physik und Astronomie,
Universit\"at Basel, Klingelbergstr. 82, CH--4056 Basel,
Switzerland} 
\author[U.\ Weiss]{U. Weiss\inst{4}}
\address[\inst{4}]{Institut f\"ur Theoretische Physik,
Universit\"at Stuttgart, Pfaffenwaldring 57, D--70550 Stuttgart,
Germany} 
 \dedicatory{The article is dedicated to Hermann Grabert on the occasion
of his 60th birthday.}

\begin{abstract}
 We review recent advances in the field of full counting
 statistics (FCS) of charge transfer through impurities imbedded into
 strongly correlated one-dimensional metallic systems, modelled by
 Tomonaga-Luttinger liquids (TLLs). We
 concentrate on the exact analytic solutions for the cumulant
 generating function (CGF), which became available recently and apply
 these methods in order to obtain the FCS of a non-trivial contact
 between two crossed TLL.

\end{abstract}
\maketitle





\section{Introduction}
\label{sect1}

The relative abundance of exactly solvable models makes the
one-dimensional (1D) systems a unique testing ground for new concepts
in condensed matter physics. However, during the last two decades
the purely academical purpose of 1D studies transformed into an
applied science as such extremely reliable 1D metallic materials
as single wall carbon nanotubes (SWNTs) became widespread in
laboratories \cite{iijima}. The surging interest in
low-dimensional materials is motivated by the rapidly shrinking
lateral sizes of the microelectronic circuitry, which is soon
expected to arrive at device dimensions at which any wire is truly
one-dimensional from the point of view of current carrying
electrons.

The most fundamental difference between the 1D metals and their
higher-dimensional counterparts is the role of interactions. While
(at least in clean systems) the full perturbative expansion in
correlation strength is controllable and even convergent in 3D and
leads only to insignificant renormalization of parameters
(Landau's Fermi liquid theory), it diverges for one-dimensional
systems. This is a clear indication of a new universality class --
the so-called Tomonaga-Luttinger liquid (TLL). In its simplest
form it has first been discussed by Tomonaga in \cite{tomonaga}. A
slight generalization was offered by Luttinger \cite{luttinger}.
Finally, an extremely convenient field theoretical reformulation
has been presented by Haldane \cite{haldane}.

It turns out, that whatever the precise form of the interaction
potential, the generic correlation term in the TLL universality
class is the short ranged $\delta$-shaped interaction. This and
the reduced dimensionality allows one to diagonalize the
corresponding Hamiltonian with minimal effort just by a canonical
transformation. The physical properties of such systems are very
interesting. For instance, there is an algebraic singularity in
the single particle density of states around the Fermi energy
$E_F$ due to a formation of highly correlated collective state.
The exponent of the corresponding power law is related to the
interaction constant. This effect is reflected in the transport
properties of such 1D systems contacted by bulk 3D electrodes -
the differential conductance is a power law of the applied voltage
or temperature whatever is larger \cite{kf,furusakinagaosa}. This
has been seen in a number of experiments as a zero-bias anomaly
\cite{bockrath,yao}.

The existence of exact solutions makes the correlated 1D metals
extremely interesting for the emerging field of full counting
statistics (FCS) \cite{nazarov}. The number of non-interacting
electrons which are transferred through a structureless scatterer
during a certain waiting time is expected to obey the binomial
distribution \cite{levitovlesovik}. It is unlikely that it is
still valid as soon as the interaction potential among electrons
becomes non-zero. In fact, it is not even clear that under such
conditions genuine physical electrons are involved in the
transport processes. For instance, in fractional quantum Hall
(FQH) samples the currents are believed to be carried by
fractionally charged Laughlin quasiparticles which have been
observed in noise experiments \cite{depicciotto,saminadayar}.
\footnote{It has been shown in
  Ref.~\cite{Tra04} that the analysis of the data in Ref.~\cite{depicciotto}
  is indeed compatible with the TLL theory of FQH edges states.} Therefore
the study of such systems would give important insights into the
non-equilibrium properties of strongly correlated electronic
systems.

Here we would like to review the developments in the field of FCS
of such strongly correlated 1D systems. We start with a brief
description of the generic setup, whose transport characteristics
one is usually interested in in Section \ref{sect2}. After that,
in Section \ref{sect3}, we present an exact analytic solution for
the charge transfer statistics, which is possible at one special
interaction strength. A generic way to solve the problem at
arbitrary interaction strength is discussed in the subsequent
Section \ref{sect4}. Finally, as an application of the developed
formalism we present some new results for the FCS of a contact
between two TLLs in Section \ref{sect5}.

\section{Impurity in a Tomonaga-Luttinger liquid}
\label{sect2}

The starting point are the free 1D electrons with a linearized
dispersion relation. As a result, the kinetic part describes two
different fermion species: the right/left-movers,
\[
 H_0 = i v_F \int d x \, \left[ R^\dag(x) \partial_x R(x) -
 L^\dag(x) \partial_x L(x) \right] \, ,
\]
where $R(x)$ and $L(x)$ are the corresponding field operators,
$\hbar =1$, and $v_F$ is the Fermi velocity. As one is mostly
interested in the system's properties in the universality region
which is essentially a low-energy regime, the linearization of the
dispersion is unproblematic. Such electron segregation is even
natural in FQH systems, where the R/L moving branches are
spatially separated from each other and are not even allowed to
interact in any way. The generic interaction of the Luttinger
liquid universality class turns out to be extremely short ranged
and effectively given by a $\delta$-shaped potential, the
corresponding Hamiltonian contribution being
\[
 H_I = U \int \, d x d y \, \rho(x) \delta(x-y) \rho(y) \, ,
\]
where $U$ is the interaction strength and $\rho(x)$ is the full
particle density operator. It turns out that the fastest way to a
diagonal Hamiltonian leads through the bosonization
representation, where both the kinetic as well as the interaction
terms have similar shapes and can be merged by a mere rotation. In
this way one obtains
\begin{eqnarray}             \label{HB}
 H_B = \frac{1}{4 \pi g} \int d x \, \left[ (\partial_x \phi_R)^2
 + (\partial_x \phi_L)^2 \right] \, ,
 \end{eqnarray}
where $\phi_{R,L}(x)$ are the bosonic fields which can be directly
related to the current density operators $j_{R/L}(x) = \mp
\partial_x \phi_{R/L}(x)$. Remarkably, the interaction strength is
now reworked into the dimensionless parameter $g = 1/\sqrt{1 +
U/(\pi v_F)}$, which in the case of an FQH bar is obtained
differently to be equal to the filling factor. To come that far it
is usually not necessary to know the precise form of the relation
between the phase fields and the original physical fermions -- the
bosonization identity. However, it is necessary as soon as one
needs to generate the bosonic form of the interedge tunnelling
operator $\sim R^\dag(0) L(0) + L^\dag(0) R(0)$. Fortunately, in
the continuum limit this relation reduces up to an overall
numerical prefactor to a rather simple prescription $R/L(0)
\rightarrow \exp\left[ i \phi_{R/L}(0) \right]$
\cite{mattis,book}. As at least in the FQHE set-up the forward
scattering can clearly be dropped from the outset\footnote{Of
course, in case of the conventional TLL the forward scattering is
always present. It can easily be taken care of by a mere phase
shift of the bosonic phases though.}, the full Hamiltonian then
reads $H = H_B + H_{BS}$, where the back scattering (in the case
of chiral TLL) and interedge tunnelling (in the case of a FQH
device) contribution is given by
\begin{equation}\label{HB1}
 H_{BS} = \lambda_{BS} \cos \left[ \phi_L(0) - \phi_R(0) \right]
 \, .
\end{equation}
$\lambda_{BS}$ is the (bare) back scattering amplitude. $H$ is
equivalent to the so-called boundary sine-Gordon (BSG)
Hamiltonian. From the renormalization group point of view it is
known to possess two fixed points: an infrared (low energy) and
ultraviolet (high energy) ones. In the infrared limit, which is
the one we are interested in, the system becomes strongly coupled
for $0<g<1$ (repulsive interactions) and $\lambda_{BS}$ flows to
infinity. In the opposite situation $g>1$ (attractive
interactions) $\lambda_{BS}$ is irrelevant, we do not consider
this situation in the rest of the paper.

\section{Analytic solution via refermionization}
\label{sect3}

As has been realized by Guinea in another context, the sine-Gordon
Hamiltonian is simply diagonalizable not only at the trivial point
$g=1$, but also at $g=1/2$ as well \cite{guinea}. Indeed,
$\exp\left[ i \phi_{L/R} (x) \right]$ trivially satisfies the
original fermionic anticommutation relation at $g=1$. On the other
hand, setting $g=1/2$ one finds that the construction $\exp\left\{
i \left[ \phi_{R} (x) - \phi_L(-x) \right] \right\}$ supplemented
by appropriate prefactors is subject to fermionic statistics as
well. In this section, we review the results of the FCS of a TLL
with a single impurity at the specific interaction parameter
$g=1/2$. The results apply as well to the FCS of the FQH interedge
tunnelling as the problems are isomorph.

In the theory of FCS, the quantity of interest is the generating
function $\chi(\lambda)$, which is the Fourier transform of the
probability distribution $P(Q)$ of the charge $Q$ crossing the
impurity during time $\tau$, $\chi(\lambda) = \sum_Q e^{i \lambda
Q}_{} P(Q)$. The function $\chi(\lambda)$ generates moments of the
charge $Q_\tau = \int_0^\tau dt I(t)$ transferred during time
$\tau$,
\begin{equation} \label{gen_fun}
\chi(\lambda) = \sum_k \frac{(i \lambda)^k}{k!} \langle Q_\tau^k
\rangle = \exp \left[ \tau \sum_k \frac{(i \lambda)^k}{k!} \langle
\delta^k Q \rangle \right] ,
\end{equation}
where $\tau  \langle \delta^k Q \rangle$ is the $k$th cumulant of the
distribution and $I(t)$ is the time-dependent current through the scattering region.
The most powerful tool to calculate the resulting $\chi$ is the Keldysh
formalism. It has been shown in Ref.~\cite{KT} that the generating function
for the single impurity problem at $g=1/2$ may be written as
\begin{equation} \label{Ham}
\chi(\lambda)= \chi_0(\lambda)\left\langle {\cal T}_{\pm} \,e^{{i}
\int_0^{\tau}{dt\, {H}^-(t)}} \,e^{-{i} \int_0^{\tau}{dt\, {H}^+(t)}} \right\rangle
\end{equation}
with time-dependent Hamiltonians
\begin{equation} \label{hpm}
H^{\pm}(t)= H_B + \int dx \,   2\lambda_{BS} \delta(x) \cos
\left\{ 2[\phi_L(0) - \phi_R(0)] \mp \frac{1}{4} \lambda -
\frac{1}{2} e \tilde{V}t\right\} ,
\end{equation}
where ${\cal T}_{\pm}$ orders operators along the Keldysh contour and
\begin{equation}
\chi_0(\lambda) = \exp\left[\tau \frac{e^2 }{4\pi}(i V \lambda - T
  \lambda^2)\right]
\end{equation}
is the generating function in the absence of backscattering.
Furthermore, $\tilde{V} = V + 2i \lambda T$, where $V$ is the
applied voltage (that drops at the impurity site), and $T$ is the
temperature. If we follow Matveev \cite{Mat95}, we can transform
the Hamiltonian (\ref{hpm}) in an elegant way and express it in
terms of a new fermion. This makes the calculation of the
generating function a straightforward exercise with the resulting
expression
\begin{eqnarray} \label{stat}
\ln{\chi}(\lambda) &=& \ln{\chi}_0(\lambda) \\
&+& \tau \int_0^{\infty}{
  \frac{d\omega}{2\pi}\, \ln\Bigl\{1+\frac{T_B^2}{\omega^2+T_B^2} }
\left[ (e^{-i \lambda}-1)\tilde{f}^+(1-\tilde{f}^-)+ (e^{i
\lambda}-1)\tilde{f}^-(1-\tilde{f}^+)\right]\Bigr\} \nonumber
\end{eqnarray}
with $\tilde{f}^+=\tilde{f}(\omega) =
\{\exp[(\omega-e\tilde{V}/2)/T]+1\}^{-1}$,
$\tilde{f}^-(\omega)=1-\tilde{f}(-\omega)$, and $T_B = a
\lambda_{BS}^2/2 v_F$ ($a^{-1}$ is the high-energy cutoff of the
TLL theory. $a$ is of the order of the lattice spacing of
the underlying lattice model.).
It has been shown in App.~C of Ref.~\cite{AndersonFCS} that
Eq.~(\ref{stat}) can be brought into the more familiar form
\begin{eqnarray} \label{stat2}
\ln{\chi}(\lambda) = \tau \int_0^{\infty}{
  \frac{d\omega}{2\pi}\, \ln\Bigl\{1+\frac{\omega^2}{\omega^2+T_B^2} }
\left[ (e^{i \lambda}-1)f^+(1-f^-)+ (e^{-i
\lambda}-1)f^-(1-f^+)\right]\Bigr\}
\end{eqnarray}
with $f^+=f(\omega) = \{\exp[(\omega-eV/2)/T]+1\}^{-1}$ and
$f^-(\omega)=1-f(-\omega)$. The two equivalent ways of writing the
generating function Eqs.~(\ref{stat}) and (\ref{stat2}) are dual
to each other. The reason is that Eq.~(\ref{stat}) employs a
formulation of the generating function using the energy-dependent
reflection coefficient ${\cal R}(\omega) = T_B^2/(\omega^2+T_B^2)$
whereas Eq.~(\ref{stat2}) employs a formulation of the generating
function using the energy-dependent transmission coefficient
${\cal T}(\omega) = \omega^2/(\omega^2+T_B^2)$. In Ref.~\cite{KT},
the first three cumulants have been calculated in terms of special
functions for arbitrary transmission, temperature, and applied
bias\footnote{We have to keep in mind that all
  energy scales should be much lower than the high-energy cutoff of the model
  $1/a$.}. They read ($G_0=e^2/h$ is the conductance quantum)
\begin{eqnarray}
\langle \delta Q \rangle &=&  \frac{1}{2} G_0 \, V \left[1-{2
T_B\over e V} \;
  {\rm Im} \; \psi\left({1\over 2}+
{2 T_B + i e V\over 4\pi T}\right)\right], \label{c1exakt} \\
\langle \delta^2 Q \rangle &=& 2 T{d \langle \delta Q \rangle \over dV} - \frac{e}{2} T_B
\coth\left({eV\over 2 T}\right) {d \langle \delta Q \rangle \over dT_B} + T  T_B{d^2 \langle \delta Q \rangle \over dVdT_B}, \label{c2exakt} \\
\langle \delta^3 Q \rangle &=& \frac{1}{2} T T_B \frac{d^2 \langle \delta^2 Q
  \rangle}{dV dT_B}
- \frac{e}{2} T_B \coth\left(\frac{eV}{2T}\right) \frac{d \langle \delta^2 Q
  \rangle}{dT_B} + 2T \frac{d \langle \delta^2 Q \rangle}{dV} \nonumber \\
&+&
\frac{e}{2} T T_B \coth\left(\frac{eV}{2T} \right)
\frac{d^2 \langle \delta Q \rangle}{dVdT_B} +
 \frac{e^2}{4} T_B \sinh^{-2}\left(\frac{eV}{2T} \right) \frac{d \langle
  \delta Q \rangle}{dT_B} ,  \label{c3exakt}
\end{eqnarray}
where $\psi$ is the digamma function. In the zero temperature
limit, Eqs.~(\ref{c1exakt}) - (\ref{c3exakt}) obey the relations
(\ref{cumrel}) between the higher order cumulants and $\langle
\delta Q \rangle$ derived earlier in Ref.~\cite{saleurweiss}.

\section{Thermodynamic Bethe ansatz solution at arbitrary interaction}
\label{sect4}

As has already been mentioned above, the system can be mapped onto
the sine-Gordon model by a folding and a subsequent transformation
to the even-odd basis \cite{ghoshal,FLS,FLSPRL}. After that
procedure the excitations in the system are easily identifiable by
means of the Bethe ansatz technique: at integer $1/g$ there are
current carrying charged kinks-antikinks, which we from now on
denote by subscripts $\pm$, and $j= 1 \dots 1/g-2$ neutral (charge
zero) breathers, which are bound states of kinks and antikinks.
Both excitation types are masseless and possess simple dispersion
relations in terms of rapidity $\theta$: $e_\pm = e^\theta/2$ and
$e_j = \sin \left[ \pi j /2(1/g-1) \right] \, e^\theta$. It is
important to realise that these excitations are not related in any
simple way to the original electrons. Nevertheless, only
(anti)kinks are scattered on the boundary and with a known quite
simple scattering matrix $S_{ij}$ given in Ref.~\cite{ghoshal}.
For the latter we shall only need the resulting effective
``transmission coefficient'' of the (anti)kinks through the
boundary,
\begin{eqnarray}                 \label{transmcoeff}
 {\cal T}(\theta) = \left| S_{++}(\theta) \right|^2 = \left| S_{--}(\theta) \right|^2
 = \frac{1}{1 + e^{2(1-1/g)(\theta - \theta_{BS})}} \, ,
\end{eqnarray}
where $\theta_{BS}$ is the effective rapidity corresponding to the
back scattering amplitude $\lambda_{BS}$. In the non-interacting
case the knowledge of the impurity transmission and of the energy
distribution function of the incoming particles (essentially Fermi
distributions) would be sufficient for construction of not only
the non-linear $I-V$ but of the CGF as well \cite{levitovlesovik}.
This is different in the present case as not only are the
quasiparticles distributed in a completely different way, the
distribution functions of the (anti)kinks [$\eta_\mp(k)$, $k$
being the particle momenta] at different energies (or rapidities)
are correlated due to interactions in the bulk. However, that does
not matter in the expression for the current through the system.
Just as in the non-interacting situation the corresponding
expression only involves linear combinations of Fermi
distributions and it is natural that no higher correlations of
population probabilities emerge even in the interacting case. In
fact, as is shown in \cite{FLS,FLSPRL,FS}, one obtains for the
current
\begin{eqnarray}    \label{IV}
 I = I(V,T_{BS}) = \frac{1}{L} \sum_k \, {\cal T}(k) (\eta_+ - \eta_-) = \int
 d \theta {\cal T}(\theta) \left[ \bar{P}_+(\theta) -
 \bar{P}_-(\theta) \right] \, ,
\end{eqnarray}
where $\bar{P}_\pm(\theta)$ is the expectation value of the
density of the occupied states, $T_{BS}$ is the temperature
(energy) scale associated with $\lambda_{BS}$ and $L$ is the
system length which is sent to infinity in the continuum limit (s.
below). This situation changes completely as soon as one goes over
to the noise properties of the system. Here even in the
non-interacting case products of Fermi distribution functions
emerge. Thus one has to expect some more involved bilinear
products of $\eta_\pm$ at different momenta to enter the
expression. This indeed has been shown in \cite{FS}. The physical
interpretation of the emerging picture is then very appealing: the
transport is mediated by one by one elastic scattering of
(anti)kinks off the boundary with the known scattering matrix. If
only one channel with particles with momenta $k$ were involved the
corresponding momentum generating function would be given by
\begin{eqnarray}            \label{smallchi}
 \chi_k(\lambda) = \left\{ 1 + {\cal T}(k) \left[ \eta_+ (1-\eta_-) (e^{-i
 \lambda} - 1) + \eta_- (1-\eta_+) (e^{ i
 \lambda} - 1) \right] \right\}^{\tau/L} \, .
\end{eqnarray}
$\tau$ is again the waiting time and $1/L$ corresponds to the
spectral weight of the single $k$-channel. It can easily be shown
that for transport in multiple channels (at different energies)
the corresponding momentum generating function is a product of
single channel contributions $\hat{\chi}(\lambda) = \prod_k \,
\chi_k (\lambda)$. The role of the breathers in the bulk however,
is to entangle the single-particle energy distribution functions
of the (anti)kinks, hence the previous formula has to be averaged
over all possible system configurations. Therefore we arrive at
the following total CGF,
\begin{eqnarray}                 \label{startpoint}
 \ln \,  \chi(\lambda) = \ln \langle \hat{\chi}(\lambda) \rangle
 = \ln \left\langle \prod_k
 \left\{ 1 + {\cal T}(k) \left[ (e^{i \lambda} - 1) \eta_- (1-\eta_+) + (e^{-i \lambda} - 1)
 \eta_+
 (1-\eta_-)\right] \right\}^{\tau/L} \right\rangle
\end{eqnarray}
which was suggested in \cite{komniksaleur}.

The exact results for $g=1/2$ follow from this ansatz quite
naturally. First of all under these conditions there are no
breathers in the bulk and the (anti)kinks are free fermions. That
is why they fill the momentum space like ordinary electrons and
are subject to Fermi distributions $\eta_\pm(k) = n_{L/R}(k) =
n_F(k \mp V/2)$. Since the quasiparticles are now non-interacting,
the averaging in (\ref{startpoint}) is superfluous and must be
dropped. Furthermore, the transmission coefficient
(\ref{transmcoeff}) is given by ${\cal T}(k) = k^2/(k^2 +
\lambda_{BS}^2)$ as $k= D e^\theta$ and $\lambda_{BS} = D
e^{\theta_{BS}}$, $D$ being a cut-off which falls out of any
observable. Combining everything we immediately obtain the result
(\ref{stat2}).

Another important limiting case is the zero temperature regime.
Here all correlations among the distributions $\eta_\pm(k)$ vanish
and all irreducible momenta turn out to be given by
\cite{saleurweiss}
\[
 \langle \delta^n Q \rangle = \frac{\tau}{L} \sum_k \, F_n(k) (\eta_+ - \eta_-) = \tau \int
 d \theta F_n(\theta) \left[ \bar{P}_+(\theta) -
 \bar{P}_-(\theta) \right] \, ,
\]
where $F_1 (\theta) = {\cal T}(\theta)$, $F_2 (\theta) = {\cal
T}(\theta)[1 - {\cal T}(\theta)]$, $F_3 (\theta) = {\cal
T}(\theta) [1 - {\cal T}(\theta)][1 - 2 {\cal T}(\theta)]$ etc. A
following very convenient relation between different $F_n$-functions can be verified,
\[
 F_n(\theta) = \frac{\partial^{n-1}}{(\partial
 \widetilde{\theta}_{BS})^{n-1}} \, F_1(\theta) \, ,
\]
with $\widetilde{\theta}_{BS} = - 2 (1 - g) g \theta_{BS}$, which
helps to reduce every single cumulant to a calculation of the
current derivative with respect to impurity rapidity,
\begin{equation}\label{cumrel}
\langle \delta^{n}_{} Q\rangle\,=\, \frac{\partial^{n-1}_{}}{(\partial\widetilde\theta_{BS}^{})^{n-1}_{}} \langle \delta Q\rangle \; .
\end{equation}

Summing up all cumulants into the generating functional one obtains
\[
 \frac{\partial}{(\partial
 \widetilde{\theta}_{BS})} \ln \chi (\lambda) = \sum_{m=1}^\infty
 \frac{(i \lambda)^m}{m\!} \frac{\partial^m}{(\partial
 \widetilde{\theta}_{BS})^m} \, I(V,\widetilde{\theta}_{BS}) =
 I(V,\widetilde{\theta}_{BS}+ i \lambda) -
 I(V,\widetilde{\theta}_{BS})\, .
\]
Although in general no closed analytic expressions for the current
exist, the expansions for strong and weak back scattering turn out
to be very useful. There are clear physical pictures in these limits \cite{saleurweiss}.
When the system is in vicinity of the weak coupling point, the true ground state is that of two
completely disconnected systems. It is evident that then only physical electrons with integer charge may tunnel between the
subsystems.
On the other hand, in the strongly coupled system a collective state between the edges is formed, which in the case of
a FQH bar has fractionally charged Laughlin quasiparticles for elementary excitations.

For strong backscattering or weak tunneling, $V/T'_{BS} \ll 1$, where $T'_B$ is defined by
\begin{equation}
T'_B/T_B = (2 \sqrt{\pi}/g) \Gamma\{ 1/2
+ g/[2(1-g)]\}/\Gamma\{ g/[2 (g-1)]\} \, ,
\end{equation}
one obtains
\begin{eqnarray}      \label{wc1}
 \ln \chi(\lambda) = \tau \sum_{m=1}^\infty \, (e^{i \lambda m}_{} - 1) k_m^{} \, ,
\end{eqnarray}
where
\begin{equation}\label{wc2}
k_m^{} \,=\, \frac{1}{m}\, a_m^{}(1/g)\, G_0^{} V (V/T'_{BS})^{2 m (1/g - 1)}
\end{equation}
with \cite{FLS}
\begin{equation}\label{wc3}
a_m^{}(g) \,=\, (-1)^{m+1} \, \frac{\Gamma(3/2) \Gamma(m g)}{
\Gamma(m) \, \Gamma[3/2 + m (g-1)]} \, .
\end{equation}
Observe that the counting field $\lambda$ enters the expression
(\ref{wc1}) with an integer prefactor $m$.\footnote{It has been
argued in \cite{levitovreznikov} that in the general case the CGF
contains prefactors $\exp( \pm i \lambda |e^*/e| )$, where $e^*$
is the charge of current carrying excitations.} The physical
meaning of the expression (\ref{wc1}) is quite illuminating.
Suppose that $k_m^{}$ is the probability per unit time to transfer
a particle of charge $m$ through the impurity barrier. Then the
charge transferred in the time interval $\tau$ is the result of a
Poisson process for particles of charge one crossing the barrier,
contributing a current $I_1^{} = k_1^{}$, plus a Poisson process
for particles of charge two contributing a current $I_2^{} =
2k_2^{}$, etc. The logarithm of the Fourier transform of the
probability distribution of all these Poisson processes would then
be the expression (\ref{wc1}). The only subtle point now is that
the signs of the rates $k_m^{}$ are not quite right since a
classical process would require that all rates are positive.
Certainly, the $m=1$--term is indeed a Poisson process for the
tunneling of single electrons,  but the joint tunneling of pairs
of electrons (and of multiples thereof) $m=2,\,4,\cdots$ comes
with the different sign, which is an effect of quantum
interference.

In the opposite limit of weak backscattering,
the expression for $\ln\chi(\lambda)$ reads
\begin{eqnarray}      \label{wb1}
 \ln \chi(\lambda) = \tau\left[ i\lambda\, g\, G_0^{}V +  \sum_{m=1}^\infty \,
 (e^{-i \lambda g\,m}_{} - 1) \widetilde k_m^{} \right]\, ,
\end{eqnarray}
where
\begin{equation}\label{wb2}
\widetilde k_m^{} \,=\, \frac{1}{m}\, a_m^{}(g)\, g\, G_0^{} V (V/T'_{BS})^{2 m (g - 1)}    \;.
\end{equation}
This form is quite similar to (\ref{wc1}), but there are subtle differences.
The first term in the expression (\ref{wb1}) represents the current of fractionally charged quasiparticles
in the absence of the barrier. The exponential factor $e^{-i\lambda g\, m}_{}$ indicates that now we have tunneling of quasiparticles
of charge $g$ and of multiples thereof, and the minus sign in the exponent means that the tunneling diminishes the current
instead of building it up as in the strong-backscattering regime. The sign of the tunneling rate $\widetilde k_m^{}$ is now that of
$\cos(m\pi g )$. Therefore, the perception of clusters of quasiparticles with fractional charge $g$ tunneling independently with
a classical Poisson process is quite appropriate when $g$ is small. The classical limit is reached as $g$ goes to zero. Then
all the rates $\widetilde k_m^{}$ are positive, but the quantum fluctuations
$\left\langle \delta^n Q \right\rangle$ with $n>1$ have faded away,
\begin{eqnarray}      \label{wb3}
 \ln \chi(\lambda)\; =\; i \lambda \tau\,g\, G_0^{}
 V \left[ 1 - \sum_{m=1}^\infty \frac{\Gamma ( m- {\textstyle\frac{1}{2}} ) }{2\sqrt{\pi} m!}
\Big(\frac{T'_{BS}}{V}\Big)^{2 m}\right] \;=\; i \lambda\tau\, g\,
G_0^{} V \sqrt{1 - \Big(\frac{T'_{BS}}{V}\Big)^{2} }   \, .
\end{eqnarray}
These results have been derived relying on the integrable
approach. Alternatively, one may return to the bosonic
representation, the BSG model (\ref{HB}) with (\ref{HB1}). Upon
integrating out the Luttinger liquid modes away from the impurity
located at $x=0$, one arrives at the Coulomb gas representation of
the quantum impurity problem which at zero temperature involves
long-range logarithmic interactions. We see from this
representation that the Luttinger liquid modes act as an Ohmic
thermal reservoir \cite{wb99}. Reconsidering the cumulant
generating function in the Coulomb gas representation with the
rigorous nonequilibrium Keldysh or Feynman-Vernon method, one
finds confirmation of the expressions (\ref{wc1}) -- (\ref{wb2})
and deeper understanding of the cumulant relation (\ref{cumrel})
\cite{saleurweiss,bfw}.

For arbitrary impurity strength or finite temperatures one has to
evaluate (\ref{startpoint}) with the help of thermodynamic Bethe
ansatz technique. The evaluation of $\eta_i(k)$ products is rather
demanding. In most of the experiments, however, still only the
cumulants of lowest orders are measured. In \cite{komniksaleur} a
general procedure for the calculation of cumulants of any order is
presented and the third irreducible moment is explicitly
calculated.

Interestingly, the leading low-temperature contribution to the zero temperature cumulant generating function can be calculated
in analytic form. To this aim, we work in the Coulomb gas representation  and perform the low temperature expansion
of the charge interaction factor, which is a power series in $T^2_{}$. We find that the $T^2_{}$--contribution to the
cumulant generating function
can be universally written  in terms of the second derivative of the zero temperature expression with respect to the voltage, whereas
higher-order terms are non-universal. In the end we find the CGF for strong backscattering in the form (\ref{wc1}) with the rate expressions
$k_m^{}$ given by
\begin{equation}\label{lt1}
k_m^{}(V,T) =  \Big[ 1 + \frac{\pi^2}{3g}
T^2_{}\frac{\partial^2}{(\partial V)^2}\Big] k_m^{}(V)
  =  \Big\{ 1 + \frac{\pi^2}{3g}
{\textstyle (\frac{2}{g}-2) m \big[1+(\frac{2}{g}-2)m\big] }
\frac{T^2}{V^2}\Big\} \, k_m^{}(V) \,,
\end{equation}
where terms of order $T^4$ are disregarded. In the opposite weak-backscattering limit we obtain in generalization
of the expression (\ref{wb1})
\begin{eqnarray}      \label{lt2}
 \ln \chi(\lambda) = \tau\left[ \big(i\lambda\,V  - \lambda^2_{}T\big) g\, G_0^{}  +  \sum_{m=1}^\infty \,
(e^{-i \lambda g\,m}_{} - 1) \widetilde k_m^{}(V,T) \right]\, ,
\end{eqnarray}
where (modulo terms of order $T^4$)
\begin{equation}
\widetilde k_m^{}(V,T) =  \Big[ 1 + \frac{\pi^2 g}{3}
T^2_{}\frac{\partial^2}{(\partial\, gV)^2}\Big] \, k_m^{}(V)
  =  \Big\{ 1 + \frac{\pi^2 }{3g}
 ( 2g-2) m \big[1+(2g-2)m\big]  \frac{T^2}{V^2}\Big\} \, \widetilde k_m^{}(V) \;.
\end{equation}
The $T^2$-contribution to the cumulants is a distinctive signature of interacting 1D electrons.
A related phenomenon is the universal $T^2_{}$-behavior observed at low temperatures in open quantum systems with Ohmic dissipation,
e.g. $T^2_{}$-enhancement in macroscopic quantum tunneling \cite{wb99}. It is also the origin of the universal Wilson ratio,
observed e.g. in Kondo systems.

\section{Application: FCS of the crossed TLLs set-up}
\label{sect5}

It is expected that as the lateral dimension of the
microelectronic circuitry becomes smaller, the ultimate nanoscale
wiring would become truly one-dimensional from the point of view
of current carrying electrons. Indeed, a metallic wire with a
thickness of only several nm can only support electrons
propagating as an $s$-wave since the states with higher angular
momenta lie so high that even at room temperature they cannot be
excited. Thus, in the future nanoelectronics one has strongly
interacting quantum wires as one of the basic circuitry elements.
The next to a simple wire complexity level is achieved by a point
contact of two such conductors. In fact, following the theoretical
proposal of \cite{XLLPRL} a number of successful experiments has
been already conducted on SWNT junctions measuring the
corresponding non-linear $I-V$ \cite{kim,gao}. We expect that
noise or even FCS measurement will be conducted on such set-ups in
the nearest future.

The most general way of coupling two conductors is not only by a
particle exchange via tunnelling, but by an electrostatic
interaction as well. As has been discussed in \cite{XLLPRL}, in
the true fixed point the inter-wire tunnelling turns out to be
completely irrelevant. Moreover, although the density-density
Coulomb coupling between the wires consists of a number of terms
only one of them with the lowest scaling dimension survives in the
fixed point. The effective Hamiltonian is then given by
\begin{eqnarray}
 H = \sum_{j=1,2} H_{B j}[\phi_{R/L j}] + \lambda_C
 \sin \left[ \phi_{L 1} (0) - \phi_{R 1}(0) \right] \,
 \sin \left[ \phi_{L 2} (0) - \phi_{R 2}(0) \right] \, ,
\end{eqnarray}
where $H_{B j}[\phi_{R/L j}]$ is given by (\ref{HB}) for the wire
$j=1,2$ and $\lambda_C$ is the coupling constant. A very
convenient simplification of the problem is achieved by
introduction of channel (anti)symmetric variables according to the
prescription\footnote{Here the equality of the interaction
constants $g_1=g_2=g$ is assumed, but this requirement is not
restrictive in any way and a more general transformation can
easily be conceived.}
\begin{eqnarray}
 \phi_{R/L \pm} = \frac{\phi_{R/L 1} \pm \phi_{R/L 2}}{\sqrt{2}}
 \, .
\end{eqnarray}
In this way the original problem separates into two commuting
contributions,
\begin{eqnarray}
 H = \sum_{p=\pm} H_{B p}[\phi_{R/L p}] + p \, \frac{\lambda_C}{2}
 \cos \left\{\sqrt{2} [\phi_{L p} (0) - \phi_{R p}(0)] \right\} \,
 .
\end{eqnarray}
It turns out that by the above rotation even the non-equilibrium
transport problem separates. Not only the voltage applied to the
new $\pm$ channels is given by the (anti)symmetric combination of
the original $V_{1,2}$, but the physical currents in the $1,2$
channels are calculated via prescription $I_{1,2} = (I_+ \pm
I_-)/\sqrt{2}$. In fact, this separation is even more fundamental:
it is easily realised that even the FCS should factorize into two
independent contributions with counting fields $\lambda_\pm =
(\lambda_1 \pm \lambda_2)/\sqrt{2}$ as
\[
 \ln {\chi} (\lambda_1, \lambda_2) = \ln \left[ \chi \left(\lambda_+, V_+\right) \, \chi
 \left(\lambda_-, V_- \right) \right] \, ,
\]
where $\chi(\lambda,V)$ is given by (\ref{startpoint}). Of course,
the first order cumulant with respect to either of the channels
leads to the non-linear $I-V$ as discussed in \cite{XLLPRL},
provided one employs the ``voltage drop'' way of coupling to the
voltage sources \cite{kf}.

Now we want to analyse correlations of higher orders. The direct
(within one of the channels $j=1,2$) noise contribution is
trivially found by double derivative,
\begin{eqnarray}
 \langle \delta^2 Q_{j} \rangle = (-i)^2
 \frac{\partial^2}{\partial \lambda^2_{j}} \, \ln {\chi} (\lambda_1, \lambda_2)
 = \left( \frac{\partial \lambda_-}{\partial \lambda_i}
 \right)^2 \, P_- + \left( \frac{\partial \lambda_+}{\partial \lambda_i}
 \right)^2 \, P_+ \, ,
\end{eqnarray}
where
\[
 P_\pm = - \frac{ \chi(\lambda_\pm,V_\pm) \, \partial_\pm^2
 \chi(\lambda_\pm,V_\pm)- [ \partial_\pm  \chi(\lambda_\pm,V_\pm)]^2}{
 \chi^2(\lambda_\pm,V_\pm)} \, .
\]
The corresponding explicit formulas are given in \cite{XLLnoise}.
As $(\partial \lambda_\pm /\partial \lambda_{1,2})^2 = 1/2$ the
noise locking found in \cite{XLLnoise} follows  immediately,
\begin{eqnarray}     \label{locking}
 \langle \delta^2 Q_{j} \rangle = (P_+ + P_-)/2 \, .
\end{eqnarray}
Another interesting quantity is the yet unstudied cross
correlation
\begin{eqnarray}   \label{finresult}
 \langle \delta Q_{1} \delta Q_2 \rangle &=& (-i)^2
 \frac{\partial^2}{\partial \lambda_{1} \partial \lambda_2} \,
 \ln {\chi} (\lambda_1, \lambda_2)
 = \left( \frac{\partial \lambda_-}{\partial \lambda_1}
 \right) \left( \frac{\partial \lambda_-}{\partial \lambda_2}
 \right)\, P_- + \left( \frac{\partial \lambda_+}{\partial \lambda_1}
 \right)\left( \frac{\partial \lambda_+}{\partial \lambda_2}
 \right) \, P_+
 \nonumber \\
 &=& (P_+ - P_-)/2 \, .
\end{eqnarray}
Remarkably, this correlation vanishes completely when there is no
net transport in either of the wires. Even more surprising feature
is that this kind of cancellation is temperature-independent. Thus
the correlations (\ref{finresult}) are unaffected by the
Johnson-Nyquist noise. This phenomenon along with the effect
(\ref{locking}) could be used in future experiments to identify
the sophisticated multi-particle state the system is in.

\section{Conclusions}

In most cases the task of FCS calculation in a generic system
requires application of nonequilibrium techniques. Usually this is
a rather challenging endeavour for strongly correlated systems,
where pronounced deviations from the conventional charge transfer
statistics are expected. One class of such systems covers
interedge tunneling in the fractional quantum Hall bars and
impurities imbedded into chiral Tomonaga-Luttinger liquids. Here
the charge of current carrying excitations is different from the
elementary charge. This results in a unique FCS, which differs
considerably from the noninteracting setups.

Despite the fact, that these systems are integrable with help of
the thermodynamic Bethe ansatz, there is no possibility to exploit
it directly to address their FCS. Nevertheless, it still can be
accessed using a number of shortcuts. One of them is the
possibility to map the corresponding boundary sine-Gordon model to
a noninteracting fermionic theory via refermionization procedure
at some special nontrivial interaction strength. The resulting
Hamiltonian turns out to be quadratic and easily diagonalizable
with elementary methods making a genuine nonequilibrium evaluation
of the FCS a rather simple task.

Another way towards the FCS is offered by the equilibrium Bethe
ansatz solution, which is able to identify all possible
excitations in the system as well as the corresponding scattering
matrices. It turns out that for the generic interaction strength
there are only two excitations which are responsible for the
transport: kinks and antikinks, whose reflections on the boundary
correspond to the physical current carrying processes. All other
excitations are merely responsible for the rather involved
(non-Fermi like) energy distribution functions. The values for the
probabilities to occupy (anti)kink states at different momenta
$\eta_\pm(k)$ are not mutually correlated as long as the system is
at zero temperature. Thus the application of elementary
statistical arguments immediately leads to the FCS. The same
result could be achieved by a straightforward Coulomb gas
expansion method. Interestingly, this approach is also able to
supply the leading corrections around the $T=0$ result.

At finite temperatures the distributions $\eta_\pm(k)$ are
correlated with each other prohibiting the FCS calculation along
this path. The cumulant generating function $\chi(\lambda)$ can be
evaluated nonetheless by an ansatz which takes advantage of the
fact, that the current transport is mediated by momentum
conserving scattering of (anti)kinks off the boundary. Indeed, the
FCS can be expressed as a complicated average of a product of
$\chi_k(\lambda)$s describing transport in individual channels
corresponding to a designated momentum value.

All these techniques are not restricted to the single impurity
problems only. They can rather be applied to access the FCS of
more complicated junctions and networks of quantum wires in the
TLL state. We expect, that this kind of information is especially
interesting in view of recent advances in the fields of
nanoelectronics, spintronics and quantum computation.

\begin{acknowledgement}
AK acknowledges the support by the DFG grant KO 2235/2. BT was
supported by the Swiss NSF and the NCCR Nanoscience. The authors
would like to thank Hubert Saleur and Markus Kindermann for
numerous fruitful discussions.
\end{acknowledgement}


\begin{thebibliography}{10}

\bibitem{iijima} S.~Iijima, Nature \textbf{354}, 56 (1991).

\bibitem{tomonaga} S.~Tomonaga, Prog.~Theor.~Phys. Osaka \textbf{5}, 349 (1950).

\bibitem{luttinger} J.~M.~Luttinger, J.~Math.~Phys. \textbf{4}, 1154 (1963).

\bibitem{haldane} F.~D.~M.~Haldane, J.~Phys.~C: Solid State Phys.
\textbf{14}, 2585 (1981).

\bibitem{kf} C.~L.~Kane and M.~P.~A.~Fisher, Phys.~Rev.~B
\textbf{46}, 15233 (1992).

\bibitem{furusakinagaosa} A.~Furusaki and N.~Nagaosa, Phys.~Rev.~B
\textbf{47}, 4631 (1993).

\bibitem{bockrath} M.~Bockrath \emph{et al.}, Nature \textbf{397},
598 (1999).

\bibitem{yao} Z.~Yao \emph{et al.}, Nature \textbf{402}, 273
(1999).

\bibitem{nazarov} Yu.~V.~Nazarov, see this volume.

\bibitem{levitovlesovik} L.~S.~Levitov and G.~B.~Lesovik, JETP
lett. \textbf{58}, 230 (1993).

\bibitem{depicciotto} R.~de~Picciotto \emph{et al.}, Nature
\textbf{389}, 162 (1997).

\bibitem{saminadayar} L.~Saminadayar \emph{et al.},
Phys.~Rev.~Lett. \textbf{79}, 2526 (1997).

\bibitem{Tra04}
B.~Trauzettel, P.~Roche, D.~C.~Glattli, and H.~Saleur, Phys.~Rev.~B
\textbf{70}, 233301 (2004).

\bibitem{mattis} D.~C.~Mattis, J.~Math.~Phys. \textbf{15}, 609
(1974).

\bibitem{book} A.~O.~Gogolin, A.~A.~Nersesyan, and A.~M.~Tsvelick,
\emph{Bosonization and strongly correlated systems}, Cambridge
Unviersity Press (1998).

\bibitem{guinea} F.~Guinea, Phys.~Rev.~B \textbf{32}, 7518 (1985);
see also U.~Weiss \emph{et al.}, Phys.~Rev.~B \textbf{52}, 16707
(1995).

\bibitem{KT} M.~Kindermann and B.~Trauzettel, Phys.~Rev.~Lett.
\textbf{94}, 166803 (2005).

\bibitem{Mat95}
K.~A.~Matveev, Phys.~Rev.~B \textbf{51}, 1743 (1995).

\bibitem{AndersonFCS} A.~O.~Gogolin and A.~Komnik, Phys.~Rev.~B
\textbf{73}, 195301 (2006).

\bibitem{ghoshal} S.~Ghoshal and A.~B.~Zamolodchikov,
Int.~J.~Mod.~Phys. A, \textbf{9}, 3841 (1994).

\bibitem{FLS} P.~Fendley, A.~W.~W.~Ludwig, and H.~Saleur,
Phys.~Rev.~B \textbf{52}, 8934 (1995).

\bibitem{FLSPRL} P.~Fendley, A.~W.~W.~Ludwig, and H.~Saleur,
Phys.~Rev.~Lett. \textbf{74}, 3005 (1995).

\bibitem{FS} P.~Fendley and H.~Saleur,
Phys.~Rev.~B \textbf{54}, 10845 (1996).

\bibitem{komniksaleur} A.~Komnik and H.~Saleur,
Phys.~Rev.~Lett. \textbf{96}, 216406 (2006).

\bibitem{saleurweiss} H.~Saleur and U.~Weiss,
Phys.~Rev.~B \textbf{63}, 201302R (2001).

\bibitem{levitovreznikov} L.~S.~Levitov and M.~Reznikov,
Phys.~Rev.~B \textbf{70}, 115305 (2004).

\bibitem{bfw} H.~Baur, A.~Fubini, and U.~Weiss, Phys. Rev. B \textbf{70},
024302 (2004).

\bibitem {wb99} U. Weiss, {\em Quantum Dissipative Systems}, 2nd ed. (World Scientific, Singapore, 1999).

\bibitem{XLLPRL} A.~Komnik and R.~Egger, Phys.~Rev.~Lett.
\textbf{80}, 2881 (1998).

\bibitem{kim} J.~Kim, K.~Kang, J.-O.~Lee, K.-H.~Yoo, J.-R.~Kim,
J.~W.~Park, H.~M.~So, and J.~J.~Kim, J.~Phys.~Soc.~Jpn.
\textbf{70}, 1464 (2001).

\bibitem{gao} B.~Gao, A.~Komnik, R.~Egger, D.~C.~Glattli, and
A.~Bachtold, Phys.~Rev.~Lett. \textbf{92}, 216804 (2004).

\bibitem{XLLnoise} B.~Trauzettel, R.~Egger, and H.~Grabert,
Phys.~Rev.~Lett. \textbf{88}, 116401 (2002).


\end{thebibliography}
\end{document}